\documentclass[12pt]{article}
\usepackage{axodraw}
\usepackage{epsfig}
\def\e{{\rm e}}
\newcommand{\be}{\begin{equation}}
\newcommand{\ee}{\end{equation}}
\newcommand{\bea}{\begin{eqnarray}}
\newcommand{\eea}{\end{eqnarray}}

\newcommand{\gm}{\gamma}
\newcommand{\Gm}{\Gamma}

\newcommand{\ep}{\epsilon}

\newcommand{\lm}{\lambda}

\newcommand{\dd}{\mbox{d}}

\newcommand{\nn}{\nonumber}

\newcommand{\Li}[2]{{\mbox{Li}}_{#1}\left(#2\right)}
\newcommand{\cdo}{\!\cdot\!}

\begin{document}
\parindent=1.5pc

\begin{titlepage}
\rightline{hep-ph/0209177}
\rightline{September 2002}
\bigskip
\begin{center}
{{\bf Analytical Evaluation of Double Boxes\footnote{
Talk presented at the 36th  Annual  Winter  School  on  Nuclear
and Particle Physics (Repino, Russia, 25~February -- 03 March  2002).
}} \\
\vglue 5pt
\vglue 1.0cm
{ {\large V.A. Smirnov\footnote{E-mail: smirnov@theory.sinp.msu.ru.
Supported by the Russian Foundation for Basic
Research through project 01-02-16171, INTAS through grant 00-00313,
and the Volkswagen Foundation, contract
No.~I/77788.}
} }\\
\baselineskip=14pt
\vspace{2mm}
{\em Nuclear Physics Institute of Moscow State University}\\
{\em Moscow 119899, Russia}
\vglue 0.8cm
{Abstract}}
\end{center}
\vglue 0.3cm
{\rightskip=3pc
 \leftskip=3pc
\noindent
Recent results on the analytical evaluation of double-box Feynman integrals
and the corresponding methods of evaluation are briefly reviewed.
\vglue 0.8cm}
\end{titlepage}

{\bf Introduction.}

\vspace{0.1cm}

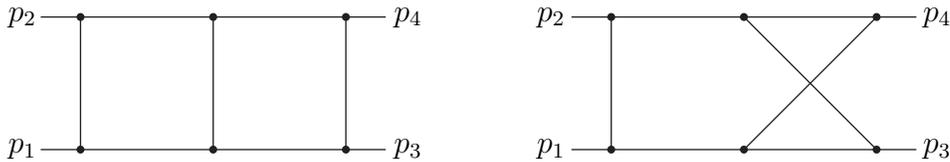
\begin {figure}[htbp]
\begin{picture}(400,100)(-50,-20)
\Line(-15,0)(0,0)
\Line(-15,50)(0,50)
\Line(115,0)(100,0)
\Line(115,50)(100,50)
\Line(0,0)(50,0)
\Line(50,0)(100,0)
\Line(100,50)(50,50)
\Line(50,50)(0,50)

\Line(0,50)(0,0)
\Line(50,0)(50,50)
\Line(100,0)(100,50)
\Vertex(0,0){1.5}
\Vertex(50,0){1.5}
\Vertex(100,0){1.5}
\Vertex(0,50){1.5}
\Vertex(50,50){1.5}
\Vertex(100,50){1.5}

\Text(-22,0)[]{$p_1$}
\Text(124,0)[]{$p_3$}
\Text(-22,50)[]{$p_2$}
\Text(124,50)[]{$p_4$}


\Line(185,0)(200,0)
\Line(185,50)(200,50)
\Line(315,0)(300,0)
\Line(315,50)(300,50)

\Line(200,0)(250,0)
\Line(250,0)(300,0)
\Line(300,50)(250,50)
\Line(250,50)(200,50)

\Line(200,50)(200,0)
\Line(250,0)(300,50)
\Line(300,0)(250,50)

\Vertex(200,0){1.5}
\Vertex(250,0){1.5}
\Vertex(300,0){1.5}
\Vertex(200,50){1.5}
\Vertex(250,50){1.5}
\Vertex(300,50){1.5}

\Text(178,0)[]{$p_1$}
\Text(324,0)[]{$p_3$}
\Text(178,50)[]{$p_2$}
\Text(324,50)[]{$p_4$}

\end{picture}
\caption{The planar and non-planar double boxes}
\label{2boxes}
\end{figure}

 {}Feynman diagrams with four external lines contribute to many important
physical quantities. They are rather complicated mathematical objects
because they are functions of multiple variables:
internal masses, two independent Mandelstam variables
$s=(p_1+p_2)^2, \;  t=(p_1+p_3)^2$ and
squares of external momenta.
The most complicated diagrams are the planar double box and
non-planar (crossed) double box shown in Fig.~\ref{2boxes}.

Almost all available analytical results correspond to the massless diagrams.
Ironically, the first result for the massless
double boxes was obtained in the most complicated case, where
all the fours external legs are off-shell, i.e.
$p_i^2\neq 0$ for all $i=1,2,3,4$. This is the elegant analytical result
for the scalar planar master double box, i.e. for all powers
of propagators equal to one, obtained in \cite{UD}:
\be
\frac{\left(i \pi^{2}\right)^2}{s^2 t}
C(p_1^2 p_4^2,p_2^2 p_3^2, s t) \,
\ee
where
\bea
C(x_1,x_2,x_3)&=& \frac{1}{\lm} \left(
6\left[ \mbox{Li}_4 (-\rho x) + \mbox{Li}_4 (-\rho y)\right]
\right.
\nn \\ && 
+3 \ln \frac{y}{x}
\left[ \mbox{Li}_3 (-\rho x) - \mbox{Li}_3 (-\rho y)\right]
+\frac{1}{2} \ln^2 \frac{y}{x}
\left[ \mbox{Li}_2 (-\rho x) + \mbox{Li}_2 (-\rho y)\right]
\nn \\ && 
\left.
+ \frac{1}{4}\ln^2 (\rho x)\ln^2 (\rho y)
+ \frac{\pi^2}{2}\ln (\rho x)\ln (\rho y)
+\frac{\pi^2}{12}\ln^2 \frac{y}{x}
+\frac{7\pi^4}{60}\right)
\;,
\label{Phi1}
\eea
\be
\lm(x,y) = \sqrt{(1-x-y)^2-4xy}\;, \;\;\; \rho(x,y)=\frac{2}{1-x-y+\lm(x,y)}
\;,
\label{lmrho}
\ee
$x=x_1/x_3, \; y=x_2/x_3$,
and Li$_n (z)$ is the polylogarithm \cite{Lewin}.

This diagram is convergent both in the ultraviolet and infrared
sense. (For example, if one puts a dot on some line,
an infrared divergence appears so that a regularization is needed.)
To obtain this result, the authors have exploited the technique
of Feynman parameters and Mellin--Barnes (MB) representation.
However, no other results for the pure off-shell double
boxes have been derived up to now so that the above result stays unique
in the pure off-shell category.

{}For massless double-box diagrams with at least one leg on the mass shell,
i.e. $p_i^2=0$, infrared and collinear divergences appear,
so that one introduces a regularization which is usually chosen to be
dimensional \cite{dimreg}, with the space-time dimension $d$ as a
regularization parameter.
One hardly believes that a regularized double-box diagram
can be analytically evaluated for the general value of the regularization
parameter $\ep=(4-d)/2$, and the evaluation is usually performed
in a Laurent expansion in $\ep$, typically, up to a finite part.

The problem of the evaluation of Feynman integrals
associated with a given graph according to some Feynman rules
is usually decomposed into two parts: reduction of general Feynman integrals
of this class to so-called master integrals (which cannot be simplified
further) and the evaluation of these master integrals.
A standard tool to solve the first part of this problem is
the method of integration by parts (IBP) \cite{IBP} when
one writes down identities obtained by putting to zero various
integrals of derivatives of the general integrand connected
with the given graph and tries to solve a resulting system of equations
to obtain recurrence relations that express Feynman integrals with
general integer powers of the propagators through the master integrals.

The purpose of this brief review is to characterize the status
of the analytical evaluation of double-box diagrams and describe
the corresponding techniques. We shall first deal with the pure
on-shell case, i.e. $p_i^2=0,\;i=1,2,3,4$, where this problem
was completely solved during last three years, i.e. all the master
integrals were calculated in expansion in $\ep$ and a reduction
procedure was developed both in the planar and non-planar case
\cite{K1,SV,Tausk,AGO,AGORT,GR0,ATT}.

We also describe a method based on alpha (or Feynman)
parameters and Mellin--Barnes (MB) representation used to calculate
the master integrals, starting with a much simpler example
of the on-shell box diagram. It turns out that the calculation of
the basic master double-box diagram \cite{K1}
based on a fivefold MB representation was far from being
optimal. As it has been shown in \cite{ATT} it is possible to
go through a fourfold MB representation. We shall describe here how this
can be done.

Then we turn to an intermediate massless case where one of the external
legs is on-shell and the other three legs are on-shell,
i.e. $p_1^2=q^2\neq 0$, $p_i^2=0,\;i=2,3,4$ and
list results of last two years where the problem
of the evaluation was solved in this situation \cite{S2,GR2}.
We then  consider the massive on-shell case and describe the only
available and recently obtained result for the master planar
double box \cite{S3}.
We conclude with a discussion of perspectives of
the analytical evaluation of the double-box diagrams.

\vspace{0.3cm}

{\bf The pure on-shell case: a box.}

\vspace{0.1cm}

\begin {figure} [htbp]
\begin{picture}(400,100)
\ArrowLine(80,0)(100,0)
\ArrowLine(80,100)(100,100)
\ArrowLine(220,0)(200,0)
\ArrowLine(220,100)(200,100)
\Line(80,0)(220,0)
\Line(80,100)(220,100)
\Line(100,0)(100,100)
\Line(200,0)(200,100)
\Text(67,0)[l]{$p_2$}
\Text(67,100)[l]{$p_1$}
\Text(228,0)[l]{$p_3$}
\Text(228,100)[l]{$p_4$}
\end{picture}
\vspace*{2mm}\\
\caption{Box diagram.}
\label{Box}
\end{figure}
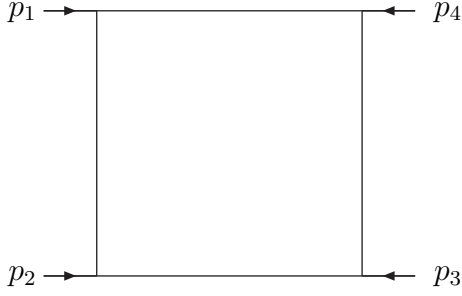

To illustrate the method of MB representation let us evaluate
the massless scalar on-shell box diagram of Fig.~\ref{Box}.
\vspace{0.5cm}
%
The Feynman integral can be written as
\be
 F(s,t;d)  =
\int\frac{\dd^dk}{(k^2+2 p_1 \cdo k) (k^2-2 p_2\cdo  k)
k^2 (k+p_1+p_2)^2}
\; ,
\label{box1}
\ee
where usual prescriptions $k^2=k^2+i 0$, etc. are implied.
Using alpha or Feynman parameters we arrive at the following
three-dimensional parametric integral:
\bea
 F(s,t;d) &=&i \pi^{d/2} \Gm(2+\ep)
\int_0^1 \int_0^1 \int_0^1 \dd \xi_1\, \dd  \xi_2 \,\dd \eta \;
\eta (1-\eta)
\nn \\  && 
\times \left[
-s \xi_1 (1- \xi_1) (1-\eta)^2
-t \xi_2 (1- \xi_2) \eta^2 -i  0
\right]^{-2-\ep} \;.
\label{box1-Fp}
\eea

Let us now use the MB representation
\be
\frac{1}{(X+Y)^{\lm}} = \frac{1}{\Gm(\lm)}
\frac{1}{2\pi  i }\int_{- i  \infty}^{+ i  \infty} \dd z
\frac{Y^z}{X^{\lm+z}} \Gm(\lm+z) \Gm(-z) \; ,
\label{MB}
\ee
where the contour of integration is chosen in the standard way:
the poles with a $\Gm(\ldots+z)$ dependence
are to the left of the contour and
the poles with a $\Gm(\ldots-z)$ dependence
are to the right of it.
Representation (\ref{MB}) is applied to the factor with square
brackets in (\ref{box1-Fp}).
As a result, the two terms in the square brackets
in (\ref{box1-Fp}) raised to the power $-2-\ep$
are replaced
by a product of some powers of these terms
(at the cost of introducing an extra integration), and, after evaluating
parametric integrals in terms of gamma functions, we obtain
\bea
 F(s,t;d) &=& \frac{i \pi^{d/2}}{(-s)^{2+\ep} \Gm(-2\ep)}\;
\frac{1}{2\pi i }
\int_{-i \infty}^{+i \infty} \dd z
\left( \frac{t}{s}\right)^z
\nn \\  && 
\times \Gm(2+\ep+z)\Gm(1+z)^2
\Gm(-1-\ep-z)^2 \Gm(-z) \; ,
\label{box1-MB}
\eea
where the same prescription for dealing with poles is implied,
i.e. the poles of $\Gm(2+\ep+z)\Gm(1+z)^2$ are to the left of the
integration contour and the poles of $\Gm(-1-\ep-z)^2 \Gm(-z)$
are to the right of it. In the case Re$\,\ep<0$, we can choose this
contour to be a straight line parallel to the imaginary axis, while
in the case Re$\,\ep>0$, a more complicated contour has to be
chosen. Anyway, if we put $\ep=0$, the integral becomes ill-defined
because the first pole with a $\Gm(\ldots+z)$ dependence, i.e.
at $z=-1$, and the first pole with a $\Gm(\ldots-z)$ dependence, i.e.
at $z=-1-\ep$, glue together and there is no space to satisfy the
prescriptions for the contours.

This observation shows how the poles of the Feynman integral
in $\ep$ are generated, from the point of view of MB integrals.
The next step is to pick up the pole in $\ep$ by taking a residue, e.g. at
the pole $z=-1-\ep$ (with the minus sign, of course) and shifting
the integration contour across this point.
A resulting integral can be taken over the line at $-1<$Re$\,z<0$.
There is no gluing in this integral so that the integral can be
safely expanded in a Taylor series in $\ep$. Every term of this
expansion can be integrated by closing the integration contour
to the right, taking residues at the points $z=0,1,2,\ldots$,
and summing up a resulting series.
Taking into account the terms up to $\ep^1$ and combining
them with the value of the above residue we arrive at the following
result:
\bea
 F(s,t;d) &=&
\frac{i \pi^{d/2}\e^{-\gm_{\rm E} \ep}}{s t}
\left\{
\frac{4}{\ep^2} - \left[\ln(-s)+\ln(-t) \right]\frac{2}{\ep}
\right.
+2 \ln(-s) \ln(-t) -\frac{4\pi^2}{3}
\nn \\  && \hspace*{-25mm}
\left.
+ \ep \left[
\Li{3}{-\frac{t}{s}} - 2\ln \frac{t}{s}\, \Li{2}{-\frac{t}{s}}
-\left( \ln^2 \frac{t}{s} +\pi^2 \right) \ln\left(1+\frac{t}{s} \right)
\right]
\right\}
+O(\ep^2)
\;,
\label{box1-res}
\eea
where $\gm_{\rm E}$ is the Euler constant.

\vspace{0.3cm}

{\bf  The pure on-shell case: the basic master diagram.}

\vspace{0.1cm}

Let us now consider the general on-shell planar
double box diagram of Fig.~1,
\vspace{0.5cm}
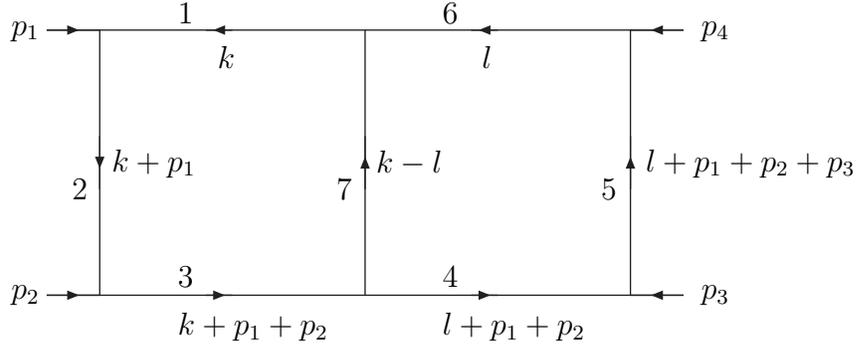
\begin {figure} [htbp]
\begin{picture}(400,100)
\ArrowLine(80,0)(100,0)
\ArrowLine(80,100)(100,100)
\ArrowLine(320,0)(300,0)
\ArrowLine(320,100)(300,100)
\Line(80,0)(320,0)
\Line(80,100)(320,100)
\Line(100,0)(100,100)
\Line(200,0)(200,100)
\Line(300,0)(300,100)
\ArrowLine(150,100)(140,100)
\ArrowLine(250,100)(240,100)
\ArrowLine(140,0)(150,0)
\ArrowLine(240,0)(250,0)
\ArrowLine(100,60)(100,40)
\ArrowLine(300,40)(300,60)
\ArrowLine(200,40)(200,60)
\Text(105,50)[l]{$ k+p_1$} %
\Text(205,50)[l]{$ k- l$}
\Text(307,50)[l]{$ l+p_1+p_2+p_3$}%
\Text(130,-12)[l]{$ k+p_1+p_2$} %
\Text(230,-12)[l]{$ l+p_1+p_2$} %
\Text(145,90)[l]{$ k$} %
\Text(245,90)[l]{$ l$}  %
\Text(230,107)[l]{6} %
\Text(130,107)[l]{1} %
\Text(230,7)[l]{4}  %
\Text(130,7)[l]{3}
\Text(90,40)[l]{2}  %
\Text(190,40)[l]{7} %
\Text(290,40)[l]{5} %
\Text(67,0)[l]{$p_2$}
\Text(67,100)[l]{$p_1$}
\Text(328,0)[l]{$p_3$}
\Text(328,100)[l]{$p_4$}
\end{picture}
\vspace*{2mm}\\
\caption{Planar double box diagram.}
\label{pdbd}
\end{figure}
%
i.e. with a general irreducible numerator and powers of propagators.
We choose  this irreducible numerator and
the routing of the external momenta as in \cite{ATT} --- see Fig.~3. For
convenience,
we consider the factor with $(k+p_1+p_2+p_3)^2$ corresponding
to the irreducible numerator
as an extra propagator but, really, we are interested only in the
non-positive integer values of $a_8$.
This general double box Feynman integral takes the form
\bea
K(a_1,\ldots,a_8;s,t;\ep) &=&
\int\int \frac{\dd^dk \, \dd^dl}{(k^2)^{a_1}[(k+p_1)^2)]^{a_2}
[(k+p_1+p_2)^2]^{a_3}}
\nn \\ && \hspace*{-50mm}
\times \frac{[(k+p_1+p_2+p_3)^2]^{-a_8}}{
[(l+p_1+p_2)^2)]^{a_4}[(l+p_1+p_2+p_3)^2]^{a_5}
(l^2)^{a_6} [(k-l)^2]^{a_7} }
\, ,
\label{2box}
\eea
where 
$k$ and $l$ are respectively loop momenta of the left and the right box.

To resolve the singularity structure of Feynman integrals in $\ep$
it is very useful to apply the MB representation (\ref{MB})
that makes it
possible to  replace sums of terms raised to some power by their
products in some powers, at the cost of introducing extra
integrations.
In \cite{K1,Tausk,S2}
MB integrations were introduced directly in alpha/Feynman
parametric integrals.
It turns out more convenient
to follow (as in \cite{ATT}) the strategy of \cite{UD} and introduce,
in a suitable way,
MB integrations, first, after integration over one of the loop momenta, $l$,
and complete this procedure after integration
over the second loop momentum, $k$.
After appropriate changes of variables we arrive at
the following fourfold MB representation of
(\ref{2box}) (see also \cite{ATT}):
\bea
K(a_1,\ldots,a_8;s,t;\ep)
 &=&
\frac{\left(i\pi^{d/2} \right)^2 (-1)^a}{
\prod_{j=2,4,5,6,7}\Gm(a_j) \Gm(4-a_{4567}-2\ep)(-s)^{a-4+2\ep}}
\nn \\ &&  \hspace*{-50mm}\times
\frac{1}{(2\pi i)^4} \int_{-i\infty}^{+i\infty}
\dd w
\dd z_2 \dd z_3 \dd z_4
\left(\frac{t}{s} \right)^{w}
\frac{\Gm(a_2 + w) \Gm(-w) \Gm(z_2 + z_4) \Gm(z_3 + z_4)}
{\Gm(a_1 + z_3 + z_4) \Gm(a_3 + z_2 + z_4)}
\nn \\ &&  \hspace*{-50mm}\times
\frac{  \Gm(4 - a_{13} - 2a_{28} - 2 \ep + z_2 + z_3)
  \Gm(a_{1238} - 2 + \ep + z_4 )\Gm(a_7 + w - z_4)
  }
{\Gm(4 - a_{46} - 2 a_{57} - 2 \ep - 2 w - z_2 - z_3)}
\nn \\ &&  \hspace*{-50mm}\times
\frac{\Gm(a_{4567} - 2 + \ep + w - z_4)
  \Gm(a_8 - z_2 - z_3 - z_4) \Gm(-w - z_2 - z_3 - z_4)}
{\Gm(4 - a_{1238} - 2 \ep + w - z_4)\Gm(a_8 - w - z_2 - z_3 - z_4) }
\nn \\ &&  \hspace*{-50mm}\times
\frac{\Gm(a_5 + w + z_2 + z_3 + z_4)
  \Gm(2 - a_{567} - \ep - w - z_2) }
  {\Gm(4 - a_{13} - 2 a_{28} - 2 \ep + z_2 + z_3 )}
  \nn \\ &&  \hspace*{-50mm}\times
\Gm(2 - a_{457} - \ep - w - z_3)
  \Gm(2 - a_{128} - \ep + z_2 ) \Gm(2 - a_{238} - \ep + z_3 )
  \nn \\ &&  \hspace*{-50mm}\times
  \Gm(4 - a_{46} - 2 a_{57} - 2 \ep - 2 w - z_2 - z_3) 
\, ,
\label{4MB}
\eea
where $a_{4567}=a_4+a_5+a_6+a_7, a_{13}=a_1+a_3$, etc., and
integration contours are chosen in the standard way.

In the case of the master double box, we set $a_i=1$ for $i=1,2,\ldots,7$
and $a_8=0$ and obtain
\bea
K^{(0)}(s,t;\ep)\equiv K(1,\ldots,1,0;s,t;\ep) &&
\nn \\ &&  \hspace*{-65mm}
= -\frac{\left(i\pi^{d/2} \right)^2}{
\Gm(-2\ep)(-s)^{3+2\ep}}
\frac{1}{(2\pi i)^4} \int_{-i\infty}^{+i\infty}
\dd w
\dd z_2 \dd z_3 \dd z_4
\left(\frac{t}{s} \right)^{w}
\frac{ \Gm(1 + w)\Gm(-w) }{\Gm(1 - 2 \ep + w - z_4) }
\nn \\ &&  \hspace*{-65mm}\times
\frac{ \Gm(2 + \ep + w - z_4)\Gm(-1 - \ep - w - z_2)
\Gm(-1 - \ep - w - z_3)}
{\Gm(1 + z_2 + z_4) \Gm(1 + z_3 + z_4)}
\nn \\ &&  \hspace*{-65mm}\times
\Gm(1 + w + z_2 + z_3 + z_4) \Gm(1 + \ep + z_4 )  \Gm(z_2 + z_4) \Gm(z_3 + z_4)
\nn \\ &&  \hspace*{-65mm}\times
\Gm(-\ep + z_2 ) \Gm(-\ep + z_3 )
\Gm(1 + w - z_4)\Gm(-z_2 - z_3 - z_4)
\, .
\label{4MB0}
\eea
Observe that, because of the presence of the factor $\Gm(-2\ep)$
in the denominator, we are forced to take some residue
in order to arrive at a non-zero result at $\ep=0$,
so that the integral is effectively threefold.

Of course, the resolution of singularities in $\ep$ in such a
multi-dimensional MB integral is much more complicated
than in the one-dimensional case. The poles in $\ep$
are not visible at once, at a first integration over one of the MB
variables. However, the rule for finding a mechanism of the generation
of poles is just a straightforward generalization of the rule used
in the previous one-loop example, where
we saw that the product of $\Gm(-1-\ep-z)\Gm(1+z)$ generated the pole
of the type $\Gm(-\ep)$ (this is nothing but the value of one of these
gamma function at the pole of the other gamma function).
Suppose we start from the integration
over one of the variables, $z$. We analyze various products
$\Gm(a+z)\Gm(b-z)$, where $a$ and $b$ depend on the rest of the
variables, with the understanding that this integration generates
a pole of the type $\Gm(a+b)$. This means that any contour of one
the next integrations should be chosen according to this dependence.
We continue this analysis with various integrations at the second
step, etc.

Here is an example of the procedure of generating poles in the integral
(\ref{4MB0}). The product $\Gm(-1 - \ep - w - z_2)\Gm(-\ep + z_2 )$
generates, due to the integration over $z_2$, a pole of the type
$\Gm(-1 - 2\ep - w)$. Then the product of this gamma function
with $\Gm(1 + w)$ generates a pole of the  type $\Gm(2\ep)$
due to the integration in $w$.

After such  preliminary analysis we conclude that the
key gamma functions that are responsible for the
generation of poles in $\ep$ are
$\Gm(-\ep + z_2 )$,  $\Gm(-\ep + z_2 )$ and $\Gm(1 + w - z_4)$.
This gives a hint for the construction of a complete procedure
of the resolution of the singularities in $\ep$, with the goal
 to decompose
the given integral into pieces where the Laurent expansion
of the integrand in $\ep$ becomes possible.
One can proceed as follows.

We first take care of the gamma functions $ \Gm(-\ep + z_2 )$ and
$ \Gm(-\ep + z_3 )$, i.e. take residues at $z_2=\ep$ and $z_3=\ep$
and shift contours across these poles. As a result, (\ref{4MB0})
is decomposed as $K=K_{11}+K_{10}+K_{01}+K_{00}$, where
$K_{11}$ corresponds to taking the two residues, $K_{00}$ is defined
by the same expression (\ref{4MB0}) but with both first poles of the
selected two gamma functions treated in the opposite way, and
the two intermediate contributions are defined by taking one of the residues
and changing the nature of the first pole of the other gamma function.

The contribution $K_{11}$ takes the form
\bea
K_{11} &=& -\frac{\left(i\pi^{d/2} \right)^2}{
\Gm(-2\ep)(-s)^{3+2\ep}}
\frac{1}{(2\pi i)^4}
\int_{-i\infty}^{+i\infty}
\dd w  \dd z_4 \left(\frac{t}{s} \right)^{w}
 \Gm(1 + w) \Gm(-1 - 2\ep - w)^2
 \nn \\ &&  
\times
\Gm(-w) \Gm(1 + w - z_4)  \Gm(2 + \ep + w - z_4)
\Gm(\ep + z_4)^2 \Gm(-2\ep - z_4)
\nn \\ &&  
\times
  \frac{\Gm(1 + 2\ep + w + z_4)}{\Gm(1 - 2\ep + w - z_4) \Gm(1 + \ep + z_4)}
\, .
\label{K11}
\eea

The contributions $K_{10}$ and $K_{01}$ are equal to each other because
of the symmetrical dependence of the integrand on $z_2$ and $z_3$.
We have
\bea
K_{01}
&=& -\frac{\left(i\pi^{d/2} \right)^2}{
\Gm(-2\ep)(-s)^{3+2\ep}}
\frac{1}{(2\pi i)^3} \int_{-i\infty}^{+i\infty}
\dd w  \dd z_2  \dd z_4
\left(\frac{t}{s} \right)^{w}
 \Gm(1 + w) \Gm(-1 - 2\ep - w)
\nn \\ &&  
\times
\Gm(-w)
 \Gm(-1 - \ep - w - z_2)\Gm(-\ep + z_2)
  \Gm(1 + w - z_4)\Gm(2 + \ep + w - z_4)
\nn \\ &&  
\frac{
\Gm(\ep + z_4)\Gm(z_2 + z_4)  \Gm(-\ep - z_2 - z_4)
  \Gm(1 + \ep + w + z_2 + z_4)}{ \Gm(1 - 2\ep + w - z_4)
\Gm(1 + z_2 + z_4)}
\, ,
\label{K01}
\eea
where the first pole of $\Gm(-\ep + z_2)$ is of the opposite nature.

{}For all these contributions, further decompositions are necessary.
One can proceed as follows.

($K_{11}$) Take minus residue at $w=-1-2\ep$. The resulting one-dimensional
integral is calculated by taking care of the first poles of
$\Gm(z_4)$ and $\Gm(z_4+\ep)$. One obtains either an explicit expression
in gamma and $\psi$ functions and a one-dimensional MB integral.
In the integral, where the first pole of $ \Gm(-1 - 2\ep - w)$
is of the opposite nature, take care of the first pole of $\Gm(z_4+\ep)$.
One obtains a one-dimensional MB integral over $w$ which is calculated
by expanding in $\ep$. The integral, where the first poles of
$ \Gm(-1 - 2\ep - w)$ and $\Gm(z_4+\ep)$ are of the opposite nature,
does not contribute because one can expand it in $\ep$ and obtain
a zero result due to $\Gm(-2\ep)$ in the denominator.

($K_{01}$) Take minus residue at $w=-1-2\ep$ and consecutively
take care of the first poles of the gamma functions
$\Gm(z_4+\ep)$, $\Gm(z_2+z_4)$ and $\Gm(z_2+z_4-\ep)$
in the resulting integral. One obtains one-dimensional MB integrals
and a two-dimensional integral
in $z_2$ and $z_4$ which is calculated by use of the second Barnes lemma
(see below).
In the integral with the pole of $ \Gm(-1 - 2\ep - w)$ of the opposite
nature, one takes care of the first pole of $\Gm(z_2+z_4)$, i.e. takes
a residue at $z_4=-z_2$ (and then takes care of the first pole of
$\Gm(-1 - \ep - w - z_2)$)
and considers an integral with the opposite
nature of this pole (with a zero contribution).

($K_{00}$) Take care of the first poles of 
$\Gm(-1 - \ep - w - z_2)$ and $\Gm(-1 - \ep - w - z_3)$.
The only non-zero contribution arises when taking both residues.

As a result we obtain either explicit expression in terms of gamma
functions and their derivatives, or one-dimensional integrals
over straight lines parallel to the imaginary axis
of ratios of gamma functions which can be of two types:
integrals over $w$ or some $z$-variable.
The integrals over $w$ can be calculated
by closing contour to the right, taking residues at the points
$w=0,1,2,\ldots$ and summing up resulting series, with the help
of the table of formulae presented in \cite{FKV}.
The one-dimensional MB integrals in $z_i$ can be calculated
with the help of
the first and the second Barnes lemmas
\bea
\frac{1}{2\pi i}\int_{-i \infty}^{+i \infty} \dd w \,
\Gm(\lm_1+w) \Gm(\lm_2+w) \Gm(\lm_3-w) \Gm(\lm_4-w)
& &
\nn \\ && \hspace*{-80mm} =
\frac{\Gm(\lm_1+\lm_3) \Gm(\lm_1+\lm_4) \Gm(\lm_2+\lm_3)
\Gm(\lm_2+\lm_4)}{\Gm(\lm_1+\lm_2+\lm_3+\lm_4)} \; ,
\label{1stBarnes}
\\
\frac{1}{2\pi i}\int_{-i \infty}^{+i \infty} \dd w \,
\frac{\Gm(\lm_1+w) \Gm(\lm_2+w) \Gm(\lm_3+w) \Gm(\lm_4-w) \Gm(\lm_5-w)}{
\Gm(\lm_1+\lm_2+\lm_3+\lm_4+\lm_5+w) }
&  &
\nn \\ && \hspace*{-120mm} =
\frac{\Gm(\lm_1+\lm_4) \Gm(\lm_2+\lm_4) \Gm(\lm_3+\lm_4)
      \Gm(\lm_1+\lm_5) \Gm(\lm_2+\lm_5) \Gm(\lm_3+\lm_5)
}{\Gm(\lm_1+\lm_2+\lm_4+\lm_5) \Gm(\lm_1+\lm_3+\lm_4+\lm_5)
\Gm(\lm_2+\lm_3+\lm_4+\lm_5)} \;
\label{2ndBarnes}
\eea
and their corollaries. These are two typical examples of such corollaries:
\bea
\frac{1}{2\pi i}\int_{-i \infty}^{+i \infty} \dd w \,
\frac{\Gm(\lm_1+w) \Gm(\lm_2+w)^2 \Gm(-\lm_2-w) \Gm(\lm_3-w)}{
\Gm(\lm_1+\lm_2+\lm_3+w) }
&  &
\nn \\ && \hspace*{-90mm} =
\frac{
\Gm(\lm_1-\lm_2) \Gm(\lm_2+\lm_3)
\left[
\psi'\left(\lm_1+\lm_3\right) - \psi'\left(\lm_2+\lm_3\right)\right]}{
\Gm(\lm_1+\lm_3)} \; ,
\label{2ndBex}
\eea
where the nature of the pole at $w=-\lm_2$ is determined
by $ \Gm(\lm_2+w)$
while other poles are treated in the standard way, and
\bea
\frac{1}{2\pi i}\int_{-1/2-i \infty}^{-1/2+i \infty} \dd w \,
\Gm(1+w) \Gm(w) \Gm(-w) \Gm(-1-w) \psi(1+w)^2
&  &
\nn \\ && \hspace*{-70mm}
=\frac{\gm_{\rm E}^2 \pi^2}{3} +6\gm_{\rm E}\zeta(3) +\frac{\pi^4}{45}
\; .
\eea
Here 
$\zeta(z)$ is the Riemann zeta function.

Collecting all the contributions, one reproduces the result of \cite{K1}:
\be
K^{(0)}(s,t;\ep)=
\frac{\left(i \pi^{d/2} \e^{-\gm_{\rm E}\ep} \right)^2 }{(-s)^{2+2\ep}(-t)}
\; K(t/s;\ep)\;,
\ee
where
\bea
K(x,\ep) &= &
-\frac{4}{\ep^4} +\frac{5\ln x}{\ep^3}
- \left( 2 \ln^2 x -\frac{5}{2} \pi^2  \right) \frac{1}{\ep^2}
\nn \\ && 
-\left( \frac{2}{3}\ln^3 x +\frac{11}{2}\pi^2 \ln x
-\frac{65}{3} \zeta(3) \right) \frac{1}{\ep}
\nn \\ && 
+\frac{4}{3}\ln^4 x +6 \pi^2 \ln^2 x
-\frac{88}{3} \zeta(3)\ln x +\frac{29}{30}\pi^4
\nn \\ && 
- \left[
2 \,\Li{3}{ -x } -2\ln x \,\Li{2}{ -x }
-\left( \ln^2 x +\pi^2 \right) \ln(1+x)
\right] \frac{2}{\ep}
\nn \\ && 
- 4 \left[S_{2,2}(-x) - \ln x\, S_{1,2}(-x)  \right]
+ 44\, \Li{4}{ -x }
\nn \\ && 
- 4 \left[\ln(1+x) + 6 \ln x  \right] \Li{3}{ -x }
\nn \\ && 
+ 2\left( \ln^2 x +2 \ln x \ln(1+x) +\frac{10}{3}\pi^2\right) \Li{2}{-x}
\nn \\ && 
+\left( \ln^2 x +\pi^2 \right) \ln^2(1+x)
\nn \\ && 
-\frac{2}{3} \left[4\ln^3 x +5\pi^2 \ln x -6\zeta(3)\right] \ln(1+x)
+ O(\ep) \;
,
\label{K1}
\eea
where, in addition to the polylogarithms, one meets also
the generalized polylogarithms  \cite{GenPolyLog}
\be
\label{Sab}
  S_{a,b}(z) = \frac{(-1)^{a+b-1}}{(a-1)! b!}
    \int_0^1 \frac{\ln^{a-1}(t)\ln^b(1-zt)}{t} \dd t \; .
\ee

This result is in agreement with the leading behaviour
in the (Regge) limit $t/s\to 0$ obtained by use of the strategy of
expansion by regions \cite{BS,SR,Sb}.
It turns out
that the (h--h), (1c--1c) and (2c--2c) contributions are the only
non-zero contributions. (See \cite{SR,Sb} for definitions of the hard (h)
and collinear ((1c) and (2c)) regions and the corresponding
contributions.) Keeping the two leading powers in $x$ we have \cite{SV}
\bea
K(x,\ep) &=&
-\frac{4}{\ep^4} +\frac{5\ln x}{\ep^3}
- \left( 2 \ln^2 x -\frac{5}{2} \pi^2  \right) \frac{1}{\ep^2}
\nn \\ &&  \hspace*{-10mm}
-\left( \frac{2}{3}\ln^3 x +\frac{11}{2}\pi^2 \ln x
-\frac{65}{3} \zeta(3) \right) \frac{1}{\ep}
\nn \\ && \hspace*{-10mm}
+\frac{4}{3}\ln^4 x +6 \pi^2 \ln^2 x
-\frac{88}{3} \zeta(3)\ln x +\frac{29}{30}\pi^4
\nn \\ && \hspace*{-10mm}
+ 2 x \left(
\frac{1}{\ep}\left(\ln^2 x - 2 \ln x +\pi^2+2\right)
\right.\nn \\ && \hspace*{-10mm}
\left.
-\frac{1}{3} \left\{
4\ln^3 x +3 \ln^2 x + (5\pi^2-36)\ln x +2 [33+5\pi^2-3\zeta(3)]
\right\}
\right)
\nn \\ && \hspace*{-10mm}
 + O(x^2 \ln^3 x ,\ep) \, .
\label{2boxNLO}
\eea

\vspace{0.3cm}

{\bf  The pure on-shell case: the non-planar case and
reduction to master diagrams.}

\vspace{0.1cm}

The basic master non-planar on-shell massless double-box diagram
were calculated in \cite{Tausk} by the same method of Feynman
parameters and MB representation, with the only qualification that
the initial integration contours were chosen, at appropriate values
of $\ep$, as straight lines parallel to imaginary axis. Then
the procedure of taking residues and shifting contours was applied,
with the requirement to keep this property.\footnote{I think, this is
a matter of taste what variant of resolution of the singularities
in $\ep$ to apply. Anyway, I have confirmed the result of
 \cite{Tausk} using the strategy described above.}
It turns out that
it is natural to consider non-planar double boxes as functions of the three
Mandelstam variables
$s,t$ and $u=(p_1+p_4)^2$ not necessarily restricted by the
physical condition $s+t+u=0$ which does not simplify the result.

Reduction procedures for the evaluation of general double-box diagrams,
with arbitrary numerators and integer powers of the propagators
were developed in \cite{SV} in the planar case and in
\cite{AGORT} in the non-planar case.
Both reduction procedures were based on solving recurrence relations
following from IBP  \cite{IBP}. In the non-planar case,
the so-called Lorentz invariance (LI) identities were
used, in addition to the IBP relations.
In \cite{SV}, the first of the two most complicated master integrals
involved is
$K^{(0)}\equiv K(1,\ldots,1,0)$ considered above. As a second
complicated master
integral, the authors of \cite{SV} have chosen the diagram with a dot
on the central line, i.e. $K(1,1,1,1,1,2,1,0)$.
As was pointed out later \cite{GT}, in practical calculations one runs into
a linear combination of these two master integrals with a coefficient
$1/\ep$, so that a problem has arisen because the calculation of
the master integrals in one more order in $\ep$ looked rather nasty.
Two solutions of this problem have immediately appeared. In \cite{GR0},
the authors have calculated this very combination
of the master integrals, while in \cite{ATT} another choice of
the master integrals has been made: instead of
$K(1,1,1,1,1,2,1,0)$, the authors have taken the integral
$K(1,1,1,1,1,1,1,-1)$ as the second complicated master integral.
This was a more successful choice because,
according to the calculational experience, no negative
powers of $\ep$ occur as coefficients at these two master integrals.

These  analytical algorithms have been successfully
applied to the evaluation of two-loop virtual
corrections to various scattering processes \cite{appl} in the
zero-mass approximation.

\vspace{0.3cm}

{\bf  One leg off-shell.}

\vspace{0.1cm}

In the case, where one of the external
legs is on-shell, $p_1^2\neq 0$, $p_i^2=0,\;i=2,3,4$,
the planar double box and one of two possible non-planar double-box
diagram with all powers of propagators equal to one
were analytically calculated in \cite{S2}, as functions of
the Mandelstam variables $s$ and $t$ and the non-zero external momentum
squared $p_1^2$. Explicit
results were expressed through  (generalized) polylogarithms, up to the
fourth order,  dependent on rational combinations of $p_1^2,s$
and~$t$, and a one- and (in the non-planar case) two-dimensional integrals
with simple integrands. To do this, the method based on MB integrals
described above was applied.

These and other master planar and non-planar double boxes with one leg
off-shell were evaluated in \cite{GR2}
with the help of the method of differential equations \cite{DE}.
The corresponding results are expressed through so-called
two-dimensional harmonic polylogarithms which
generalize harmonic polylogarithms introduced in \cite{2dHPL}.

A reduction procedure that provides the possibility
to express any given Feynman integral to the master integrals was
also developed  in \cite{GR2}. It is based on
the observation that, when increasing the total dimension of the denominator
and numerator in Feynman integrals associated with the given graph,
the total number of IBP and LI equations grows faster\footnote{This
fact was first pointed out by Laporta and used in \cite{LR}.}
than the number of independent Feynman integrals
(labeled by the powers of propagators and
powers of independent scalar products in the numerators).
Therefore this system of equations
sooner or later becomes overconstrained, and one obtains the possibility
to perform a reduction to master integrals.

In fact, this strategy is quite general and can be applied to any Feynman
graph. However its implementation for
any concrete graph looks rather non-trivial. In particular, solving
an overconstrained system of several thousand equations is a challenging
technical task.
Nevertheless, this method was successfully applied \cite{GR3}
to the Feynman
integrals with one leg off-shell contributing
to the process $e^+e^-\to 3$jets.

\vspace{0.3cm}

{\bf The massive on-shell master double box.}

\vspace{0.1cm}

\begin {figure} [htbp]
\begin{picture}(400,100)
\ArrowLine(80,0)(100,0)
\ArrowLine(80,100)(100,100)
\ArrowLine(320,0)(300,0)
\ArrowLine(320,100)(300,100)
\Line(80,0)(320,0)
\Line(80,100)(320,100)
\DashLine(100,0)(100,100){3}
\DashLine(200,0)(200,100){3}
\DashLine(300,0)(300,100){3}

\Text(67,0)[l]{$p_2$}
\Text(67,100)[l]{$p_1$}
\Text(328,0)[l]{$p_3$}
\Text(328,100)[l]{$p_4$}
\end{picture}
\vspace*{2mm}\\
\caption{Planar double box diagram. Solid and dashed lines denote
massive (with the mass $m$) and massless propagators, respectively.}
\label{2boxM}
\end{figure}
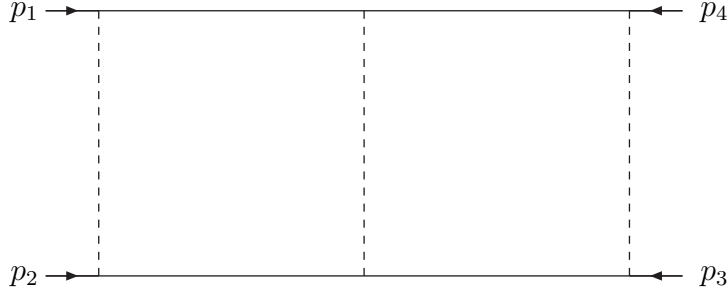

We now turn to the massive on-shell case, i.e. $p_i^2=m^2,\;i=1,2,3,4$,
and consider the general on-shell planar double box diagram
of Fig.~\ref{2boxM},
\vspace{0.5cm}
%
i.e. with general irreducible numerator and powers of propagators.
The irreducible numerator, the numbering of lines and
the routing of the external momenta are chosen as in the massless
case of (\ref{2box}) (see Fig.~\ref{pdbd}).
This general double box Feynman integral takes the form
\bea
B(a_1,\ldots,a_8;s,t,m^2;\ep) &=&
\int\int \frac{\dd^dk \, \dd^dl}{(k^2-m^2)^{a_1}[(k+p_1)^2)]^{a_2}
[(k+p_1+p_2)^2-m^2]^{a_3}}
\nn \\ && \hspace*{-50mm}
\times \frac{[(k+p_1+p_2+p_3)^2]^{-a_8}}{
[(l+p_1+p_2)^2-m^2)]^{a_4}[(l+p_1+p_2+p_3)^2]^{a_5}
(l^2-m^2)^{a_6} [(k-l)^2]^{a_7} }
\, .
\label{2box-m}
\eea
To arrive at a MB representation (with, possibly, minimal
number of MB integrations) one straightforwardly generalizes
the above procedure for the massless case
by introducing two extra MB integrations,
when separating terms with $m^2$ after each of the integrations over
the loop momenta, and after appropriate changes of variables
one obtains the following sixfold MB representation of
(\ref{2box-m}):
\bea
B(a_1,\ldots,a_8;s,t,m^2;\ep)
 &=&
\frac{\left(i\pi^{d/2} \right)^2 (-1)^a}{
\prod_{j=2,4,5,6,7}\Gm(a_j) \Gm(4-a_{4567}-2\ep)(-s)^{a-4+2\ep}}
\nn \\ &&  \hspace*{-50mm}\times
\frac{1}{(2\pi i)^6} \int_{-i\infty}^{+i\infty}
\dd w \prod_{j=1}^5 \dd z_j
\left(\frac{m^2}{-s} \right)^{z_1+z_5}
\left(\frac{t}{s} \right)^{w}
\frac{\Gm(a_2 + w) \Gm(-w) \Gm(z_2 + z_4) \Gm(z_3 + z_4)}
{\Gm(a_1 + z_3 + z_4) \Gm(a_3 + z_2 + z_4)}
\nn \\ &&  \hspace*{-50mm}\times
\frac{  \Gm(4 - a_{13} - 2a_{28} - 2 \ep + z_2 + z_3)
  \Gm(a_{1238} - 2 + \ep + z_4 + z_5)\Gm(a_7 + w - z_4)
  }
{\Gm(4 - a_{46} - 2 a_{57} - 2 \ep - 2 w - 2 z_1 - z_2 - z_3)}
\nn \\ &&  \hspace*{-50mm}\times
\frac{\Gm(a_{4567} - 2 + \ep + w + z_1 - z_4)
  \Gm(a_8 - z_2 - z_3 - z_4) \Gm(-w - z_2 - z_3 - z_4)}
{\Gm(4 - a_{1238} - 2 \ep + w - z_4)\Gm(a_8 - w - z_2 - z_3 - z_4) }
\nn \\ &&  \hspace*{-50mm}\times
\frac{\Gm(a_5 + w + z_2 + z_3 + z_4)
  \Gm(2 - a_{567} - \ep - w - z_1 - z_2) }
  {\Gm(4 - a_{13} - 2 a_{28} - 2 \ep + z_2 + z_3 - 2 z_5)}
  \nn \\ &&  \hspace*{-50mm}\times
\Gm(2 - a_{457} - \ep - w - z_1 - z_3)
  \Gm(2 - a_{128} - \ep + z_2 - z_5) \Gm(2 - a_{238} - \ep + z_3 - z_5)
  \nn \\ &&  \hspace*{-50mm}\times
  \Gm(4 - a_{46} - 2 a_{57} - 2 \ep - 2 w - z_2 - z_3) \Gm(-z_1)\Gm(-z_5)
\, .
\label{6MB}
\eea

In the case of the master double box, we set $a_i=1$ for $i=1,2,\ldots,7$
and $a_8=0$ and obtain a massive generalization of (\ref{4MB0})
\bea
B^{(0)}(s,t,m^2;\ep)\equiv B(1,\ldots,1,0;s,t,m^2;\ep) &&
\nn \\ &&  \hspace*{-75mm}
= -\frac{\left(i\pi^{d/2} \right)^2}{
\Gm(-2\ep)(-s)^{3+2\ep}}
\frac{1}{(2\pi i)^6} \int_{-i\infty}^{+i\infty}
\dd w \prod_{j=1}^5 \dd z_j
\left(\frac{m^2}{-s} \right)^{z_1+z_5}
\left(\frac{t}{s} \right)^{w}
\frac{ \Gm(1 + w)\Gm(-w) }{\Gm(1 - 2 \ep + w - z_4) }
\nn \\ &&  \hspace*{-75mm}\times
\frac{ \Gm(2 + \ep + w + z_1 - z_4)\Gm(-1 - \ep - w - z_1 - z_2)
\Gm(-1 - \ep - w - z_1 - z_3) \Gm(-z_1)}
{\Gm(1 + z_2 + z_4) \Gm(1 + z_3 + z_4)
\Gm(-2 \ep + z_2 + z_3 - 2 z_5)}
\nn \\ &&  \hspace*{-75mm}\times
\frac{\Gm(-\ep + z_2 - z_5) \Gm(-\ep + z_3 - z_5)
\Gm(1 + \ep + z_4 + z_5)\Gm(-z_5) \Gm(-2 \ep + z_2 + z_3)  }
{\Gm(-2 - 2 \ep - 2 w - 2 z_1 - z_2 - z_3)}
\nn \\ &&  \hspace*{-75mm}\times
\Gm(-2 - 2 \ep - 2 w - z_2 - z_3)
\Gm(1 + w + z_2 + z_3 + z_4)   \Gm(z_2 + z_4) \Gm(z_3 + z_4)
\nn \\ &&  \hspace*{-75mm}\times
\Gm(1 + w - z_4)\Gm(-z_2 - z_3 - z_4)
\, .
\label{6MB0}
\eea

The resolution of singularities in $\ep$ is performed
\cite{S3} also in the
standard way (see \cite{K1,Tausk,ATT,S2} and the examples above)
and reduces to shifting
contours and taking residues. The final result takes the following form:
\be
B^{(0)}(s,t,m^2;\ep) =
-\frac{\left(i\pi^{d/2}
\e^{-\gm_{\rm E}\ep} \right)^2 x^2}{s^2 (-t)^{1+2\ep}}
\left[ \frac{b_2 (x)}{\ep^2} + \frac{b_1 (x)}{\ep}
  + b_{01} (x) + b_{02} (x,y)
+ O(\ep) \right]
\label{Result}
\;,
\ee
where $x=1/\sqrt{1-4m^2/s},\; y=1/\sqrt{1-4m^2/t}$,
\bea
b_2 (x) &=&  2  (m_{x} - p_{x})^2\;,
\label{ResultEp2}
\\  
b_1 (x) &=&
 -8 \left[\Li{3}{ \frac{1 - x}{2}} + \Li{3}{\frac{1 + x}{2}} +
              \Li{3}{ \frac{-2 x}{1 - x}} + \Li{3}{\frac{2 x}{1 + x}}\right]
\nn \\ &&  
+ 4 (m_{x} - p_{x})\left[\Li{2}{\frac{1 - x}{2}}
- \Li{2}{\frac{-2 x}{1 - x}}\right] - (4/3) m_{x}^3 + 4 m_{x}^2 p_{x}
\nn \\ &&  
         - 6 m_{x} p_{x}^2 + (2/3) p_{x}^3
        + 4 l_{2} (m_{x} p_{x} + p_{x}^2)
- 2 l_{2}^2 (m_{x} + 3 p_{x})
\nn \\ &&  
- (\pi^2/3)(4 l_{2} - m_{x} - 3 p_{x}) + (8/3) l_{2}^3 + 14 \zeta_3
\;,
\label{ResultEp1}
\\
b_{01} (x) &=&
-8 (m_{x} - p_{x})\left[\Li{3}{ x} - \Li{3}{ -x} - \Li{3}{\frac{1 + x}{2}}
+ \Li{3}{\frac{1 - x}{2}}
\right.\nn \\ && \left.  \hspace*{-5mm}
- \Li{3}{\frac{2 x}{1 + x}}
+ \Li{3}{\frac{-2 x}{1 - x}}\right]
+ 4 \left[\Li{2}{ x}^2 + \Li{2}{ -x}^2 + 4 \Li{2}{\frac{1 - x}{2}}^2\right]
\nn \\ &&  
- 8  \Li{2}{ x}\Li{2}{ -x}
+ 16 \Li{2}{\frac{1 - x}{2}}(\Li{2}{ x} - \Li{2}{ -x})
\nn \\ &&  \hspace*{-15mm}
- (4/3) [\pi^2 - 6 l_{2}^2 + 3 m_{x}^2 + 6 m_{x} (2 l_{2} - 2 l_{x} - p_{x})
+ 12 l_{x} p_{x} - 3 p_{x}^2](\Li{2}{ x} - \Li{2}{ -x})
\nn \\ &&  
- (8/3) [\pi^2 - 6 l_{2}^2 + 6 l_{x} p_{x}
- 6 m_{x}( l_{x} + p_{x} - 2 l_{2})]\Li{2}{\frac{1 - x}{2}}
\nn \\ &&  
+ 8 (m_{x} - p_{x})\left[
(p_{x} - m_{x} + 2 l_{2}) \Li{2}{\frac{2 x}{1 + x}}
+ 2 (l_{x} - m_{x} + l_{2})\Li{2}{\frac{-2 x}{1 - x}}\right]
\nn \\ &&  
-8  (m_{x} - p_{x}) (2 l_{x} - p_{x} - 5 m_{x} + 4 l_{2})
(- m_{x} p_{x} + l_{2}( m_{x} + p_{x}) - l_{2}^2 + \pi^2/6)  
\nn \\ &&  
- (20/3)m_{x}^4 + (164/3)m_{x}^3 p_{x} - 40m_{x}^2 p_{x}^2
- (4/3) m_{x} p_{x}^3 - (8/3) p_{x}^4
\nn \\ &&  
+ 8m_{x} l_{x} (m_{x}^2 - 3m_{x} p_{x} + 2p_{x}^2)
\nn \\ &&  
- 4 l_{2} (7m_{x}^3 + 21m_{x}^2 p_{x} - 4m_{x} l_{x} p_{x}
- 23m_{x} p_{x}^2 + 4l_{x} p_{x}^2 - p_{x}^3)
\nn \\ &&  
- \pi^2 ((17/3)m_{x}^2 - (4/3)m_{x} l_{x} - 2m_{x} p_{x}
+ (4/3)l_{x} p_{x} - (7/3) p_{x}^2)
\nn \\ &&  
+ l_{2}^2 (84m_{x}^2 - 8m_{x} l_{x} - 16m_{x} p_{x}
+ 8l_{x} p_{x} - 44p_{x}^2)
\nn \\ &&  
- (8/3) l_{2} (6 l_{2}^2 - \pi^2) (3 m_{x} - 2 p_{x})
- (4/3)\pi^2 l_{2}^2 + 4l_{2}^4 + \pi^4/9
\;,
\label{ResultEp01}
\eea
and
\bea
b_{02} (x,y) &=&
2 ( p_{x} - m_{x})
\left\{
4\left[ \Li{3}{\frac{1 - x}{2}} - \Li{3}{\frac{1 + x}{2}}
+ \Li{3}{\frac{(1 - x) y}{1 - x y}}
\right.\right.
\nn \\ &&  \left.
- \Li{3}{\frac{-(1 + x) y}{1 - x y}} + \Li{3}{\frac{-(1 - x) y}{1 + x y}}
- \Li{3}{\frac{(1 + x) y}{1 + x y}} \right]
\nn \\ &&  
+ 2 \left[ \Li{3}{\frac{(1 + x)(1 - y)}{2(1 - x y)}}
- \Li{3}{\frac{(1 - x) (1 + y)}{2(1 - x y)}}
\right.\nn \\ &&  \left.
- \Li{3}{\frac{(1 - x) (1 - y)}{2(1 + x y)}}
+ \Li{3}{\frac{(1 + x) (1 + y)}{2(1 + x y)}} \right]
\nn \\ &&  \hspace*{-20mm}
+ 2 (m_{y} + p_{y} - m_{xy} - p_{xy})\left[2\Li{2}{ x} - 2\Li{2}{ -x}
+ \Li{2}{\frac{-2 x}{1 - x}} - \Li{2}{\frac{2 x}{1 + x}}\right]
 \nn \\ &&  
+ 4 (m_{xy} - p_{xy}) (\Li{2}{ -y} - \Li{2}{ y})
- 4 (m_{x} + p_{x} - 2l_{2}) \Li{2}{\frac{1 - x}{2}}
\nn \\ &&  
- 4 (m_{xy} - p_{xy}) \Li{2}{\frac{1 - y}{2}}
- 4 (m_{x} + l_{y} - m_{xy} )\Li{2}{\frac{(1 - x) y}{1 - x y}}
\nn \\ &&  \hspace*{-10mm}
+ 4(p_{x} + l_{y} - m_{xy})\Li{2}{\frac{ -(1 + x) y}{1 - x y}}
- 4 (m_{x} + l_{y} - p_{xy} )\Li{2}{\frac{-(1 - x) y}{1 + x y}}
\nn \\ &&  
+ 4 (p_{x} + l_{y} - p_{xy})\Li{2}{\frac{(1 + x) y}{1 + x y}}
\nn \\ &&  
+ 2 (m_{x} + p_{x} + m_{y} + p_{y} - 2m_{xy} - 2l_{2})
\Li{2}{\frac{(1 - x) (1 + y)}{2(1 - x y)}}
\nn \\ &&  
+ 2 (m_{x} + p_{x} + m_{y} + p_{y} - 2p_{xy} - 2l_{2})
\Li{2}{\frac{(1 - x) (1 - y)}{2(1 + x y)}}
\nn \\ &&  
+    
2 p_{x}^2(m_{y} + p_{y} - m_{xy} - p_{xy})
+ 2 p_{x} (2 (m_{y} l_{y} + m_{y} p_{y} + l_{y} p_{y})
\nn \\ &&  
+ m_{xy} (-m_{y} - 2l_{y} - 3p_{y} + 3m_{xy})
+ p_{xy}(-3m_{y} - 2l_{y} - p_{y} + 3p_{xy}))
\nn \\ &&  
+ 2 m_{x} (2p_{x} + m_{y} - 2l_{y} + p_{y})
(m_{y} + p_{y} - m_{xy} - p_{xy})
- p_{y}^2(m_{xy} + p_{xy})
\nn \\ &&  
+ 2p_{y}(2m_{xy}^2 + p_{xy}^2)
+ m_{y}^2(2p_{y} - m_{xy} - p_{xy})
\nn \\ &&  
+ 2m_{y}(p_{y}^2 + m_{xy}^2 + 2p_{xy}^2 - p_{y}(3m_{xy} + p_{xy}))
- 2(m_{xy}^3 + p_{xy}^3)
\nn \\ &&  
+ 2 l_{2}(
(4m_{y} + 4p_{y} - 3m_{xy})m_{xy} + (2m_{y} + 2p_{y} - 3p_{xy})p_{xy}
\nn \\ &&  
- 2(p_{x} + 2m_{x})
(m_{y} + p_{y} - m_{xy} - p_{xy}) - m_{y}^2 - 4m_{y} p_{y} - p_{y}^2)
\nn \\ &&   
+ 2l_{2}^2(3(m_{y} + p_{y}) - 2(2m_{xy} + p_{xy}))
\nn \\ && \left. 
- (\pi^2/3) (m_{y} + p_{y} - 8m_{xy} + 6p_{xy})
\right\}
\;.
\label{ResultEp02}
\eea
Here the following abbreviations are used: $\zeta_3=\zeta(3)$,
$l_z =\ln z$ for $z=x,y,2$,  $p_z =\ln(1+z)$ and
$m_z =\ln(1-z)$ for $z=x,y,xy$.

The result~(\ref{Result})--(\ref{ResultEp02})
is in agreement with the leading
power behaviour in the (Sudakov) limit
of the fixed-angle scattering, $m^2 \ll |s|, |t|$.
This asymptotics is obtained by use of the strategy of
expansion by regions  \cite{BS,SR,Sb}.
The structure of regions is very rich.
The following family of seventeen regions participates here:
\vspace{1mm}

(h--h), (1c--h), \ldots, (4c--h), (1c--1c), \ldots, (4c--4c),

(1c--3c), (2c--4c), (1c--4c), (2c--3c),

(1uc--2c), (2uc--1c), (3uc--4c), (4uc--3c).
\vspace{1mm}

\noindent
Here $h$ denotes hard, $c$ -- collinear and $uc$ -- ultracollinear
regions for the two loop momenta.
(See \cite{SR} and Chapter~8 of \cite{Sb} for definitions of these regions.)
In particular, the (h--h) contribution is nothing but the massless
on-shell double box (\ref{K1}).
Evaluating and summing up all the contributions, we obtain \cite{S3}
\bea
B^{(0)}(s,t,m^2;\ep) &=&
-\frac{\left(i\pi^{d/2}
\e^{-\gm_{\rm E}\ep} \right)^2}{s^2 (-t)^{1+2\ep}}
\left\{ 2 \frac{L^2}{\ep^2}
-\left[ (2/3) L^3+(\pi^2/3) L+2\zeta_3\right]
\frac{1}{\ep}
\right.
\nn \\ &&  
-(2/3)L^4 +2\ln(t/s) L^3
-2( \ln^2(t/s) +4\pi^2/3) L^2
\nn \\ &&  \hspace*{-37mm}
+\left[
4 \Li{3}{-t/s} -4 \ln(t/s) \Li{2}{-t/s}
+(2/3)  \ln^3(t/s)-2\ln(1+t/s)  \ln^2(t/s)
\right.\nn \\ &&  \left.\left. \hspace*{-37mm}
+(8\pi^2/3) \ln(t/s)-2\pi^2 \ln(1+t/s)+10\zeta_3
\right] L +\pi^4/36 \right\}
+ O(m^2 L^3,\ep)
\;,
\label{LO}
\eea
where $L=\ln(-m^2/s)$. This asymptotic behaviour is reproduced
when one starts from result
(\ref{Result})--(\ref{ResultEp02}).

It is interesting to note that, in the above result,
there are no so-called two-dimensional harmonic polylogarithms \cite{2dHPL}
which have turned out to be adequate functions to express results for
the double boxes with one leg off-shell \cite{GR2}.

\vspace{0.3cm}

{\bf Perspectives.}

\vspace{0.1cm}

It is believed that sooner or later we shall achieve the limit in
the process of analytical evaluation of Feynman integrals so that
we shall be forced to proceed only numerically.
(See, e.g., the first paper of \cite{Pas} where this point of view has
been emphasized.)
The progress in the field of numerical evaluation of Feynman integrals
was rather visible last years. Several new powerful numerical methods have
been developed --- see, e.g., \cite{BH,GY,Lap,Pas}.
They have been applied in practice
and, at the same time, served for crucial checks of analytical results. For
example, the method of \cite{BH} applicable to Feynman integrals with severe
ultraviolet, infrared and collinear divergences was successfully
applied to check a lot of results for the double-box master integrals
presented and/or discussed above.

However the dramatic progress in the field of
analytical evaluation of Feynman integrals shows that we
have not yet exhausted our abilities in this direction.
Let us take the problem of the analytical evaluation of the on-shell
double-box diagrams with $p_i^2=m^2,\;i=1,2,3,4$ as an example.
The calculational experience, in particular obtained
when evaluating four-point diagrams,
tell us that if such master integrals can be evaluated, the problem
can be also completely solved, after evaluating other master integrals
and constructing a recursive procedure that expresses any given Feynman
integral with general numerators and integer powers of propagators
through the master integrals. Therefore the explicit analytical result
(\ref{Result})--(\ref{ResultEp02})
can be considered as a kind of existence theorem, in the sense that it
strongly indicates the possibility to
analytically compute a general Feynman integral in this class
(and apply these results to various scattering processes in two loops
without putting masses to zero).
To calculate the master integrals one can apply the technique of
MB integration described above.
To construct appropriate recursive algorithms both in the planar and
non-planar case one can use recently
developed methods based on shifting dimension \cite{Tar} and
differential equations \cite{GR2} as well a method based on non-recursive
solutions of recurrence relations \cite{Bai} (successfully applied
in practice \cite{BCK}).

One can also hope that new analytical results can be obtained
for many other classes of Feynman integrals depending on two and three
scales. In particular, the analytical evaluation of
any two-loop two-scale Feynman integral with two, three and four legs
looks quite possible.

\end{document}